\pdfoutput=1

\documentclass[11pt]{article}

\usepackage[]{acl}

\usepackage{times}
\usepackage{latexsym}

\usepackage[T1]{fontenc}

\usepackage[utf8]{inputenc}

\usepackage{microtype}

\usepackage{amssymb}
\usepackage{amsfonts}
\usepackage{amsmath}
\usepackage{booktabs}
\usepackage{multirow}
\usepackage{graphicx}
\usepackage{enumitem}
\usepackage{textcomp}
\usepackage{booktabs}
\usepackage{subcaption}
\usepackage{xspace} 
\usepackage{diagbox}
\usepackage{amssymb} 
\usepackage{multirow}
\usepackage{xcolor}
\usepackage{xspace} 
\usepackage{color} 
\usepackage{bm}
\usepackage{colortbl}
\usepackage{setspace}
\usepackage{array}
\usepackage{arydshln} 
\usepackage{makecell}
\usepackage{xcolor,colortbl}
\usepackage{tcolorbox}
\usepackage{listings}
\usepackage{pythonhighlight}
\usepackage{wrapfig,lipsum,booktabs}
\usepackage{adjustbox}
\usepackage{natbib}
\usepackage{bbold}
\usepackage{xcolor} 
\definecolor{LightGray}{gray}{0.9}

\newcommand{\methodnamews}{\text{SRank}}
\newcommand{\methodname}{\methodnamews~}

\title{Functional Overlap Reranking for Neural Code Generation}

\author{Hung Quoc To , Minh Huynh Nguyen , Nghi D. Q. Bui \\
        FPT Software AI Center, Viet Nam \\
        \texttt{hungtq29@fpt.com, minh.nghminh@gmail.com, bdqnghi@gmail.com}}

\begin{document}
\maketitle
\begin{abstract}

Code Large Language Models (CodeLLMs) have ushered in a new era in code generation advancements. However, selecting the best code solutions from all possible CodeLLM outputs remains a challenge. Previous methods often overlooked the intricate functional similarities and interactions between solution clusters. We introduce \textit{\textbf{SRank}}, a novel reranking strategy for selecting the best solutions from code generation, focusing on modeling the relationships between clusters of solutions. By quantifying the functional overlap between solution clusters, our approach provides a better ranking strategy for code solutions. Empirical results show that our method achieves remarkable results on the pass@1 score. For instance, on the Human-Eval benchmark, we achieve 69.66\% in pass@1 with Codex002, 75.31\% with WizardCoder, 53.99\% with StarCoder, and 60.55\% with CodeGen, surpassing state-of-the-art code generation reranking methods such as CodeT and Coder-Reviewer on the same CodeLLM by a significant margin (\textbf{$\approx 6.1\%$ improvement on average}). Even in scenarios with a limited number of sampled solutions and test cases, our approach demonstrates robustness and superiority, marking a new benchmark in code generation reranking. Our implementation can be found at \url{https://github.com/FSoft-AI4Code/SRank-CodeRanker}.

\end{abstract}

\section{Introduction}

Recent advancements in language models tailored for code, known as Code Large Language Models (CodeLLMs) \citep{luo2023wizardcoder, wang2023codet5plus, nijkamp2023codegen, rozi2023codellama, wei2023magicoder, lozhkov2024starcoder, pinnaparaju2024stable, guo2024deepseek, bui2023codetf}, have garnered significant interest, particularly due to the expansion of large-scale language models and the volume of pre-training data \citep{kaplan2023scaling_laws, codegenerationwithlargelanguagemodels}. A primary utility of CodeLLMs is their capacity to generate code given natural language descriptions written by humans \citep{chen2021_eval_llms, fried2023incoder, chowdhery2022palm, nijkamp2023codegen}. However, prior studies \citep{Holtzman2020The, austin2021program_synthesis} have highlighted that the sequences generated by these models can be prone to errors, especially when likelihood-based decoding techniques like greedy search and beam search are employed. Alternatively, sampling-based decoding techniques \citep{fan-etal-2018-hierarchical, Holtzman2020The} extract multiple solutions from the model's multinomial distribution. This method generates a wide range of code solutions, many of which are correct \citep{austin2021program_synthesis, ni2023lever}. As a result, there is a growing interest in developing reranking strategies for code generation \citep{li2022competition, inala2022faultaware, chen2023codet, zhang2023coderreviewer, ni2023lever}, with the goal of sorting through an abundance of sampled solutions to identify high-quality and accurate ones.

The goal of reranking is to organize the set of candidate programs so that accurate programs are prioritized. \citet{li2022competition}, \citet{chen2023codet}, and \citet{ni2023lever} have clustered code solutions based on their functionality, then used cluster-specific data to determine ranking scores. Given that language models frequently produce code solutions that differ syntactically but are semantically analogous, functional clustering narrows the candidate pool. The emphasis then shifts from ranking individual solutions to ranking clusters themselves. Previous ranking strategies, such as AlphaCode \citep{li2022competition} and CodeT \citep{chen2023codet}, provide approaches to clustering and reranking code solutions. While AlphaCode \citep{li2022competition} focuses on identical outputs from model-generated test inputs, CodeT \citep{chen2023codet} focuses on solutions that pass model-generated test cases. The Coder-Reviewer approach \citep{zhang2023coderreviewer}, inspired by collaborative software development, uses a dual-model system to cross-check generated programs against language instructions. However, by treating clusters in isolation, they fail to model potentially informative functional similarities and interactions across clusters.

To address this limitation, we propose \textbf{\methodnamews}, a novel reranking approach emphasizing \textit{modeling inter-cluster} relationships. Specifically, we introduce a new metric called \textit{functional overlap} to quantify the similarity between clusters based on their execution outputs. This allows for identifying the most representative cluster that exhibits maximal overlap with all other clusters. As inconsistencies often indicate incorrect functionality, the cluster interacting most comprehensively likely represents the optimal solution. By incorporating these inter-cluster relationships into the ranking pipeline, we can better identify the most promising solutions. Through extensive evaluation, we demonstrate that modeling inter-cluster relationships and functional overlap provides significant and consistent improvements over prior state-of-the-art solution ranking methods \citep{li2022competition, chen2023codet, zhang2023coderreviewer} on a wide range of state-of-the-art CodeLLMs, including Codex, WizardCoder, StarCoder, and CodeGen. For instance, on the HumanEval benchmark, our method achieved a pass@1 score of 75.31\% with WizardCoder34B, outperforming the Coder-Reviewer’s score of 66.9\%. Similarly, on the MBPP-S benchmark, our method improved the pass@1 score for WizardCoder from 50.3\% with Coder-Reviewer to 51.03\% with our approach. Similar improvements are applied for other CodeLLMs, including StarCoder, CodeGen, and Codex002. If we compare \methodnamews with a simple random sampling method to get code solutions, we observe massive improvements across the models, with average improvements of 23.07\% and 17.64\% for HumanEval and MBPP-S, respectively. Our evaluation is more \textit{comprehensive} because we include many SOTA CodeLLMs of varying sizes, whereas CodeT and Coder-Reviewer did not. This provides compelling evidence of our approach's robustness across a wide range of models.

We also conducted an extensive analysis to demonstrate some of our advantages, such as our approach's remarkable robustness even with limited solutions and test cases. In summary, by moving from isolated clusters to interacting clusters with quantified functional overlap, our novel reranking strategy aims to address the limitations of prior ranking techniques for code generation.
To summarize our contributions, they are as follows:
\begin{itemize}[leftmargin=*]
    \item We introduce a novel reranking strategy for CodeLLMs that emphasizes the inter-cluster relationships and leverages the functional overlap between them, providing a more robust and accurate approach to pick the best solutions.
    \item Through extensive and comprehensive evaluations, we demonstrate that our approach consistently outperforms existing state-of-the-art methods in code generation. For instance, our method achieved superior results on both the HumanEval and MBPP-S benchmarks across various CodeLLMs.
    \item We perform extensive analysis to evaluate the robustness of our method, highlighting its effectiveness even with a limited number of sampled solutions and test cases, and its ability to capture intricate interactions between clusters, setting it apart from previous ranking techniques.
\end{itemize}

\section{Background \& Motivation}

\subsection{Code Generation}
Code generation involves generating code solutions to programming problems based on a given context $c$. The context includes a natural language description and a code snippet containing statements such as imports and a function signature. Additionally, a predefined set of test cases, denoted as $T$, is provided to evaluate the correctness of the generated code solutions. Using $c$ as the input on CodeLLM, we obtain a set of solutions $\mathbf{S} = \{s_1, s_2, ..., s_N\}$, where $N$ is a hyperparameter defining the number of return sequences from the CodeLLM execution. A solution $s$ is considered valid if it successfully passes the predefined set of test cases $\mathbf{T}$.

\begin{figure}
\centerline{\includegraphics[width=0.9\linewidth]{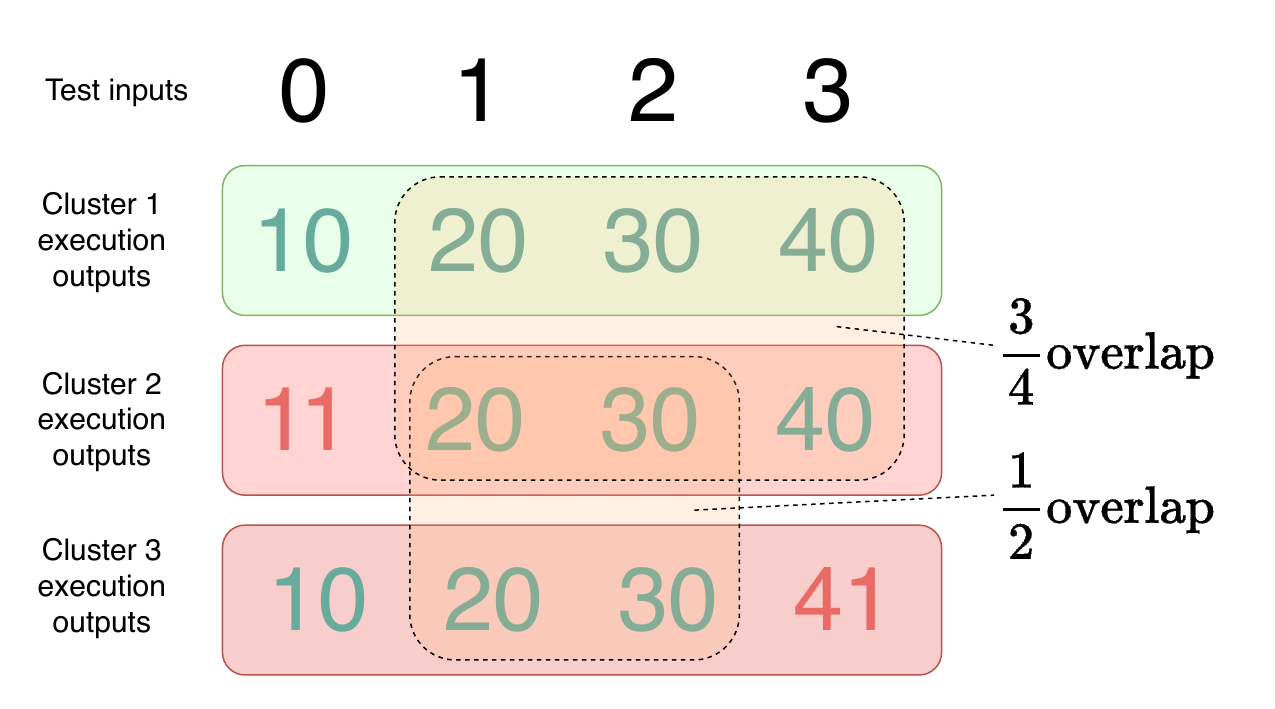}}
\caption{Illustration on concept of "functional overlap" among clusters of solutions. Cluster 1 outputs [10,20,30,40]. Cluster 2's output is [11,20,30,40]. Cluster 3's output is [10,20,30,40]. As a result, Cluster 1 overlaps Cluster 2 on three values [20,30,40], indicating that they are 3/4 overlapped. Cluster 1 overlaps Cluster 3 on three values [10,20,30], which can also be considered 3/4 overlapped. Cluster 1 has a functional overlapping score of 3 + 3 = 6. Cluster 2 overlaps with Cluster 3 on two values [20,30], resulting in a functional overlapping score of 2 + 3 = 5, and Cluster 3 has a functional overlapping score of 5. Thus, Cluster 1 has the highest cumulative functional overlap, is most representative and likely to be the optimal solution. }\label{figure:functional_overlap}
\vspace{-1em}
\end{figure}

\subsection{Solution Clustering and Reranking}
The reranking task aims to prioritize correct programs in the candidate list $\mathbf{S}$. Previously, solutions were clustered by functionality, simplifying the task due to the language models' tendency to produce syntactically varied but semantically similar solutions. Thus, the focus shifts from ranking individual solutions to ranking these functional clusters.

For instance, AlphaCode \citep{li2022competition} uses a distinct model to create test inputs. Solutions are then executed against these inputs, and those with matching outputs are clustered. The reranking strategy is based on the understanding that while there can be numerous incorrect program variations, correct ones often show similar patterns, leading to clusters being ranked by their solution count.

Conversely, CodeT \citep{chen2023codet} clusters solutions that pass the same model-generated test cases. However, this can group functionally diverse solutions. If a cluster's solutions only pass some test cases, it's uncertain if the outputs for the failed cases are consistent across solutions, potentially compromising cluster functionality and the confidence in selecting from these ranked clusters. We show a concrete example to analyze this problem in Appendix~\ref{sec:case-study}.

\subsection{Modeling Inter-Cluster Relationships}
Existing clustering and reranking methods analyze clusters independently without considering inter-cluster relationships \citep{li2022competition, chen2023codet}. However, modeling these interactions can better indicate cluster correctness. As such, we propose a new metric called \textit{``functional overlap''} to quantify cluster similarity based on execution outputs, as shown in Figure~\ref{figure:functional_overlap}. We can execute code solutions from each cluster on the same test inputs and compare their outputs. The level of output match indicates the extent of functional overlap between two clusters.

The intuition is that clusters with high overlap exhibit greater shared functionality. By modeling the extent to which a cluster overlaps with others, functional overlap identifies the most ``representative'' cluster. A cluster with maximal cumulative overlap has outputs most consistent with all other clusters. As inconsistencies often indicate incorrect functionality, the cluster interacting most comprehensively is likely the optimal solution. This is similar to the assumptions of \citet{fishler1981consensus}, where incorrect solutions are diverse and there is a low probability of having a functional agreement among incorrect solutions.

\section{Approach Details}

\subsection{Overview}

\begin{figure*}
\centerline{\includegraphics[width=1.0\linewidth]{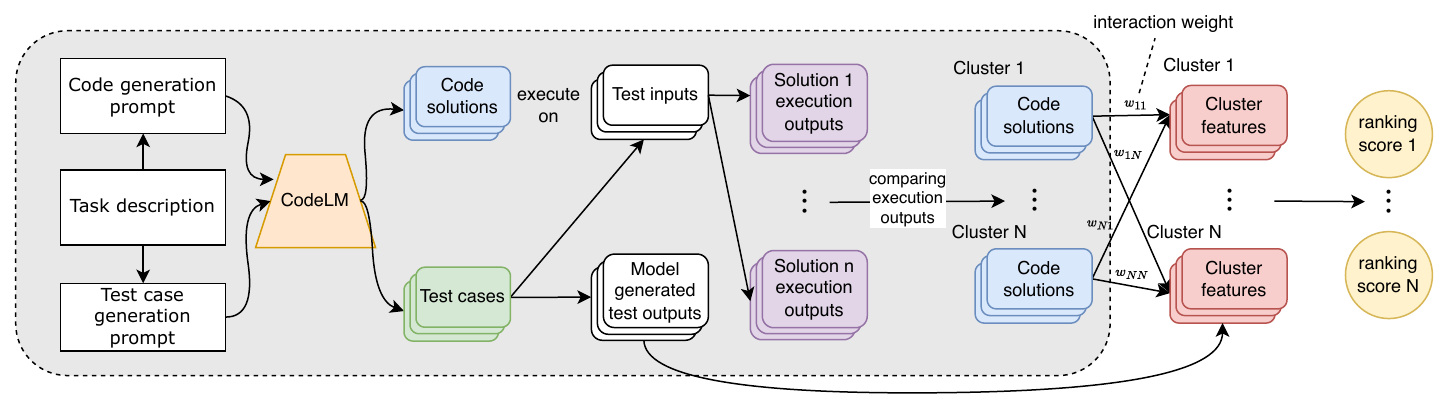}}
\caption{Method overview.}\label{figure:overview}
\vspace{-1em}
\end{figure*}

Figure \ref{figure:overview} provides an overview of our end-to-end approach. First, given a well-trained CodeLLM, e.g., Codex, and three inputs: (1) Task description, (2) Code generation prompt, (3) Test case generation prompt, we instruct the CodeLLM to generate a set of code solutions as well as test cases. Specifically, we prompt the CodeLLM to produce a collection of code solutions $\mathbf{S} = \{s_1, s_2, ..., s_N\}$ and a set of test cases $\mathbf{T} = \{t_1, t_2, ..., t_M\}$, where $N$ and $M$ are hyperparameters defining the number of solutions and test cases.

Each test case $t_i$ consists of two components: the test input $z_i$ and the expected output $\hat{o}_i$ based on the context (e.g., \texttt{assert add(1,2) == 3}, where \texttt{(1,2)} is the input and \texttt{3} is the output). We can then execute the test inputs $\mathbf{Z} = \{z_1, z_2, ..., z_M\}$ on the set of solutions $\mathbf{S}$ to generate the execution outputs $\mathbf{O} = \{o_{11}, o_{12}, ..., o_{NM}\}$.
Next, we cluster the solutions $\mathbf{S}$ into groups $\mathbf{C} = \{C_1, C_2, ..., C_K\}$ based on their execution outputs, where $K$ is the number of unique clusters. We then compute an interaction matrix $\mathbf{I}$ to quantify the functional overlap between clusters.
Finally, we multiply the interaction matrix $\mathbf{I}$ by a validation score vector $\mathbf{V}$ to obtain final ranking scores $\mathbf{R}$ for selecting the optimal solutions. The validation scores in $\mathbf{V}$ represent features of each cluster, such as the number of solutions.

In the following sections, we elaborate on the key steps of our algorithm.

\subsection{Clustering Solutions by Execution Outputs}

We first execute each solution $s_i \in \mathbf{S}$ on the test inputs $\mathbf{Z}$ to produce execution outputs $\mathbf{O}$. Solutions that exhibit identical execution outputs are grouped into the same cluster:
\[F(s_i) = F(s_j) \iff \mathbf{O}_{s_i} = \mathbf{O}_{s_j}.\]
Here, $F$ represents the clustering function that maps a solution $s$ to a cluster identifier $C_k$. The above equation indicates that two solutions $s_i$ and $s_j$ are assigned to the same cluster if and only if their output sets $\mathbf{O}_{s_i}$ and $\mathbf{O}_{s_j}$ are exactly equal.

\subsection{Computing Interaction Matrix}

After obtaining execution outputs $o_{ij}$ for each cluster $C_i$ on test input $z_j$, we define an interaction matrix $\mathbf{I} \in \mathbb{R}^{K \times K}$ to quantify functional overlap:
\begin{equation} \label{eq:interaction-matrix}
	I_{ij} = \frac{1}{M}\sum_{k=1}^{M}\delta(o_{ik} = o_{jk}),
\end{equation}
where $o_{ik}$ and $o_{jk}$ refer directly to the execution outputs of clusters $C_i$ and $C_j$, respectively, on the $k^{\text{th}}$ test input. $\delta$ is the indicator function that returns 1 if the condition inside is true and 0 otherwise.

\subsection{Computing Final Ranking Scores}

In addition to modeling inter-cluster interactions via $\mathbf{I}$, we also consider an extra validation dimension $\mathbf{V} \in \mathbb{R}^{K\times 1}$ containing cluster features. For instance, $V_i$ could represent the number of solutions in cluster $C_i$ (abbreviated as cluster sizes) or the number of test cases that the solutions in cluster $C_i$ passed (abbreviated as pass rates), providing a notion of cluster confidence. The final ranking vector $\mathbf{R} \in \mathbb{R}^{K\times 1}$ can then be computed as $\mathbf{R} = \mathbf{I} \cdot \mathbf{V}$. Here, $R_i$ aggregates information about both the inter-cluster interactions of $C_i$ (via $\mathbf{I}$) and its cluster features (via $\mathbf{V}$). Clusters with higher ranking scores in $\mathbf{R}$ are those with significant functional overlap to other clusters and high validity according to $\mathbf{V}$. By considering inter-cluster relationships and functional similarity in a principled manner, we believe our ranking approach can effectively identify the most promising solutions. We validate our method through extensive experiments in the following sections.
\section{Experimental Setup}

\paragraph{Models}
We evaluate our method on several state-of-the-art CodeLLMs, including Codex, WizardCoder, StarCoder, and CodeGen. Each model family has different model sizes (e.g., WizardCoder 15B and 34B) and different training methods (e.g., base model and instruction fine-tuned model). As a result, we chose a diverse set of models, ranging from small to large scale, and from base models to instruction fine-tuned models, to demonstrate the efficacy of our ranking method. In total, we demonstrate our approach on 6 models ranging from 6B to 34B parameters.


\paragraph{Metrics}
We use pass@k \citep{chen2021_eval_llms}, which is often employed to evaluate the functional correctness of code solutions based on code execution instead of similarity-based metrics. 

\paragraph{Baselines}
We compare SRank with recent methods for solution reranking, including Coder-Reviewer \citep{zhang2023coderreviewer} and CodeT \citep{chen2023codet}. Coder-Reviewer \citep{zhang2023coderreviewer} is the state-of-the-art method. On the other hand, CodeT \citep{chen2023codet} shares a similar clustering-reranking approach to our method.

\paragraph{Benchmarks}
We use two popular benchmarks in code generation: HumanEval \citep{chen2021_eval_llms} and MBPP-S (sanitized version) \citep{austin2021program_synthesis}. For a more challenging assessment, we evaluate \textbf{\methodname} on APPS \citep{hendrycks2021measuring}. To avoid exposing real test cases to the language model, we follow the prompt design in \cite{chen2021_eval_llms} by removing all example input-output cases from context before generating code solutions and test cases.

\paragraph{Implementation Details} 
For Codex002 and CodeGen16B, we refer to the artifacts, including both solutions and test cases, provided by \citet{chen2023codet}. Regarding the remaining models, we use the HuggingFace library \citep{wolf2019huggingface} and load models in half-precision. We set the temperature to 0.8, the top $p$ to 0.95, the maximum new tokens to 2048, and the timeout for executing solutions to 5 seconds. For each problem, we sample 100 code solutions and 100 sequences of test cases, each sequence containing multiple test cases. The prompts we used for sampling code solutions and test cases from each model can be found in Appendix~\ref{sec:appendix-prompt}. For post-processing code solutions and test cases, we follow \citet{chen2023codet} to truncate the generated sequences by the five-stop words: "\textbackslash{nclass}", "\textbackslash{ndef}", "\textbackslash{n\#}", "\textbackslash{nif}", and "\textbackslash{nprint}".

\section{Experimental Results}
\begin{table*}[t]
\begin{center}
	\scalebox{0.7}{
		\begin{tabular}{lcccccc}
			\toprule
			{} & \multicolumn{6}{c}{\textbf{HumanEval}} \\
			\cmidrule(lr){2-7}
			{} & WizardCoder34B & WizardCoder15B & CodeGen2.5-Instruct & StarCoder & Codex002 & CodeGen16B \\
			\midrule
			Greedy            & 68.90 & 50.61 & 28.05 & 39.63 & 47.00 & 29.70 \\
			CodeT             & 72.36 & 58.64 & 56.81 & 50.51 & 65.80 & 36.70 \\
			Coder-Reviewer    & - & 49.37 & 45.63 & 38.71 & 66.90 & 42.60 \\

			Random            & 59.88 & 45.20 & 26.68 & 32.55 & 37.06 & 22.78 \\
                      \midrule
			SRank  & \textbf{75.31} & \textbf{59.99} & \textbf{60.55} & \textbf{53.99} & \textbf{69.66} & \textbf{43.07} \\
			\midrule
			
			{} & \multicolumn{6}{c}{\textbf{MBPP-S}} \\
			\cmidrule(lr){2-7}
			{} & WizardCoder34B & WizardCoder15B & CodeGen2.5-Instruct & StarCoder & Codex002 & CodeGen16B \\
			\midrule
			Greedy            & 60.42 &51.29 & 42.86 & 45.90 & 58.10 & 42.40 \\
			CodeT             & 63.39 & 58.18 & 55.02 & 58.05 & 67.70 & 49.50 \\
			Coder-Reviewer    & - & 52.52 & 52.74 & 49.48 & 64.70 & 50.30 \\
			
			Random            & 54.37 & 45.72 & 34.60 & 39.26 & 47.50 & 31.54 \\
   \midrule
            SRank & \textbf{64.14} & \textbf{59.01} & \textbf{57.02} & \textbf{58.38} & \textbf{69.25} & \textbf{51.03} \\
			\bottomrule
		\end{tabular}
	}
 \captionsetup{type=table}
	\caption{Results of pass@1 on HumanEval and MBPP-S benchmarks in the zero-shot setting compared to SOTA methods, CodeT and Coder-Reviewer.}
	\label{tab:main-result}
\end{center}
\end{table*}
\begin{table*}[t]
\begin{center}
	\scalebox{0.7}{
		\begin{tabular}{lcccccc}
			\toprule
			{} & \multicolumn{6}{c}{\textbf{HumanEval}} \\
			\cmidrule(lr){2-7}
			{} & WizardCoder34B & WizardCoder15B & CodeGen2.5-Instruct & StarCoder & Codex002 & CodeGen16B \\
			\midrule
			Cluster sizes               & 72.17 & 56.38 & 55.92 & 48.63 & 59.43 & 40.51 \\
			Pass rates               & 65.09 & 43.07 & 36.17 & 35.28 & 58.37 & 21.89 \\
   Cluster sizes + Pass rates           & 73.28 & 58.21 & 58.35 & 51.90 & 66.07 & 41.72 \\
   \midrule
			Interaction + Cluster sizes      & {73.79} & {58.16} & {59.46} & {53.12} & {65.84} & {42.48} \\
			
			Interaction + Pass rates     & {73.59} & {53.49} & {48.37} & {51.13} & {65.91} & {34.61} \\
			\midrule
			SRank (all)  & \textbf{75.31} & \textbf{59.99} & \textbf{60.55} & \textbf{53.99} & \textbf{69.66} & \textbf{43.07} \\
			\midrule
			
			{} & \multicolumn{6}{c}{\textbf{MBPP-S}} \\
			\cmidrule(lr){2-7}
			{} & WizardCoder34B & WizardCoder15B & CodeGen2.5-Instruct & StarCoder & Codex002 & CodeGen16B \\
			\midrule
			
            Cluster sizes              & 65.46 & 56.17 & {56.76} & 55.50 & 64.22 & 53.00 \\
            Pass rates              & 61.88 & 49.93 & 43.80 & 48.57 & 60.80 & 36.67 \\
             Cluster sizes + Pass rates           & {64.38} & 58.13 & \textbf{57.30} & 57.68 & 68.78 & 50.60 \\
             \midrule
            Interaction + Cluster sizes     & \textbf{66.39} & {56.80} & 55.61 & {55.58} & {66.50} & \textbf{53.08} \\
			
            Interaction + Pass rates      & {63.33} & {56.11} & {50.27} & {51.33} & {64.53} & {42.78} \\
			\midrule
           
            SRank (all) & 64.14 & \textbf{59.01} & 57.02 & \textbf{58.38} & \textbf{69.25} & {51.03} \\
			\bottomrule
		\end{tabular}
	}
 \captionsetup{type=table}
	\caption{Results of Ablation Study on combining different cluster features}
	\label{tab:ablation-study}
 \end{center}
\end{table*}
\begin{table}[t]
\begin{center}
	\scalebox{0.7}{
		\begin{tabular}{lcccc}
			\toprule
			Method & Introduction & Interview & Competition \\
			\midrule
            Random            & 20.35 & 3.11 & 0.74 \\
			Greedy            & 27.20 & 5.10 & 1.80  \\
			CodeT             & 34.60 & 8.10 & 2.20 \\
                      \midrule
			SRank  & \textbf{37.79} & \textbf{9.53} & \textbf{3.29} \\
			\bottomrule
		\end{tabular}
	}
 \captionsetup{type=table}
	\caption{Results of pass@1 on APPS benchmakr using Codex002 model in the zero-shot setting compared baselines.}
	\label{tab:apps-result}
\end{center}
\end{table}


\begin{figure*}[t]
    \centering
    \includegraphics[width=1.0\linewidth]{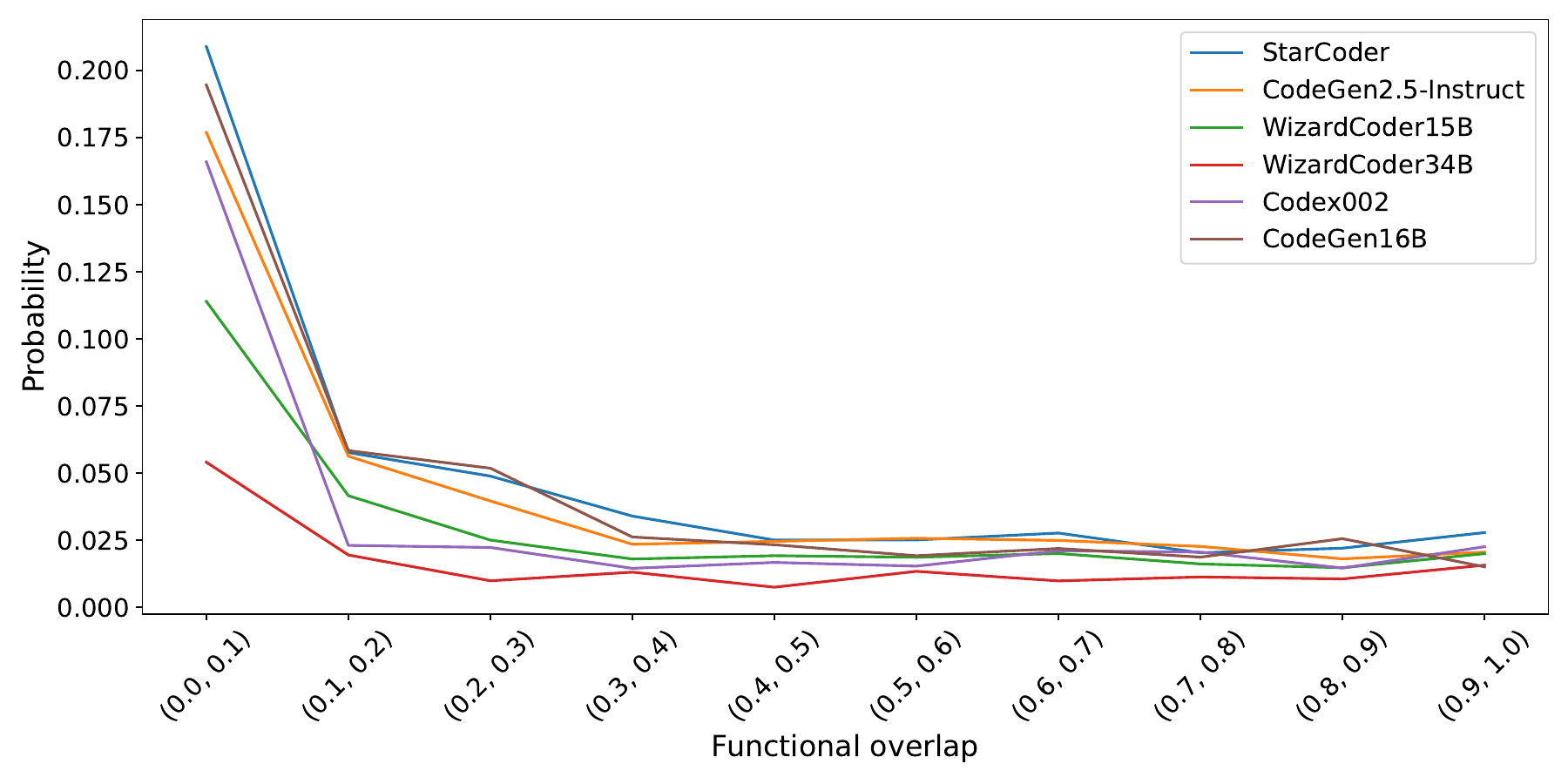}
    \caption{Probability of incorrect solutions varied based on the degree of functional agreement on HumanEval.}
    \label{fig:prob-incorrect-solutions}
\end{figure*}
Table \ref{tab:main-result} presents the pass@1 results on the HumanEval and MBPP-S benchmarks for various CodeLLMs. Our method, SRank, consistently outperforms other techniques across most models and benchmarks. For instance, on the HumanEval benchmark, {\methodnamews} achieves average improvements over CodeT and Coder-Reviewer of about 3.63\% and 8.81\% in pass@1, respectively.

Additionally, when comparing Coder-Reviewer with the random sampling method, it is unstable across the models. Specifically, using WizardCoder15B and StarCoder as examples, Coder-Reviewer brings modest increases of 4.17\% and 6.16\%, compared to our improvements of 14.79\% and 21.44\%. On the MBPP-S benchmark, \textbf{\methodnamews} still achieves outstanding performance, although the magnitude of improvements is slightly less than that of HumanEval. Our comprehensive experiments demonstrate the effectiveness of \textbf{\methodnamews} over CodeT and Coder-Reviewer.

To assess the effectiveness of our proposed method on a wide range of coding problem difficulties, we evaluate \textbf{\methodname} on APPS using the Codex002. Results are shown in Table \ref{tab:apps-result}. We observe that \textbf{\methodnamews} consistently outperforms all other baselines by a significant margin. Compared with CodeT, adding inter-cluster modeling significantly improves reranking results. On the other hand, due to the difficulty of the tasks, the improvement is less significant when the level of difficulty increases. The results show that \textbf{\methodnamews} is robust and scales well with different difficulty levels.

In addition, to validate our method on closed source and high capability models, we select Anthropic Claude 3 Opus \citep{anthropic2024claude3} as a representative LLM in our experiment. Please refer to Appendix~\ref{sec:closed-source-results} for results.
\section{Analysis}
\label{sec:analysis}

\paragraph{Assumption Validation}
We provide a comprehensive analysis to validate our assumption that \textit{incorrect solutions are diverse and there is a low probability of functional agreement among incorrect solutions}. Formally, let $\mathbf{S}$ denote the set of solutions sampled from a certain CodeLLM, $s_i$ be the $i$-th solution in $\mathbf{S}$, $\mathbf{C_k}$ be the $k$-th cluster by our clustering algorithm, $C^i$ be the cluster including the solution $s_i$, and $|\mathbf{C}_k|$ and $|\mathbf{C}_k|^{*}$ be the number of solutions and quantity of incorrect solutions in the $k$-th cluster, respectively. The function $f(s_i, s_j)$ is defined as the computation of the functional overlap between $s_i$ and $s_j$, similar to Eq. \ref{eq:interaction-matrix}. We then compute the probability of incorrect solutions with varying levels of functional overlap.
\begin{equation}
\begin{split}
    p(\text{l} \le f(s_i, s_j) < \text{h}, s_i
    ~\text{and} ~s_j~ \text{are incorrect}) \\= \frac{\sum_{(i, j) \in \mathcal{M}}{|C^i|^{*}|C^j|^{*}}}{\binom{|S|}{2}}
\end{split}
\end{equation}
Here, $\mathcal{M}$ is comprised of pairs of $(s_i, s_j)$ where $\text{l} \le f(s_i, s_j) < \text{h}$, and $\text{l}$ and $\text{h}$ are the two hyperparameters. We consider two range values, $(\text{l}_1, \text{h}_1)$ and $(\text{l}_2, \text{h}_2)$ with the same length, where $\text{l}_1 < \text{l}_2$ and $\text{h}_1 < \text{h}_2$. According to our assumption, we anticipate that the following inequality holds:
\begin{equation*}
\begin{split}
    p(\text{l}_1 \le f(s_i, s_j) < \text{h}_1, s_i
    ~\text{and} ~s_j~ \text{are incorrect}) > \\ p(\text{l}_2 \le f(s_i, s_j) < \text{h}_2, s_i 
    ~\text{and} ~s_j~ \text{are incorrect}) 
\end{split}
\end{equation*}

The left term denotes the probability of lower functional agreement among incorrect solutions, while the right term signifies the corresponding probability of higher functional agreement among incorrect solutions. The observed relationship indicates that the probability of lower functional agreement surpasses that of higher functional agreement, substantiating our assumption. Results in Figure \ref{fig:prob-incorrect-solutions} show a general decline in probability with increasing values of $\text{l}$ and $\text{h}$. Particularly for the range $(\text{l}, \text{h}) = (0, 0.1)$, the probability is significantly higher compared to those of other ranges, and at the next range, $(\text{l}, \text{h}) = (0.1, 0.2)$, the probability experiences a notable decrease. Moreover, when incorrect solutions exhibit a high functional overlap, exceeding 0.7, the probability is low at around 3\%. These findings are consistent with our assumption. More results can be found in Appendix \ref{sec:appendix-valid-assumption}.

\paragraph{Impact of Cluster Features}
We aim to evaluate reranking performance solely with cluster features, i.e., only cluster sizes, pass rates, or a combination of both. Table \ref{tab:ablation-study} shows SRank's performance when these features are added or removed from the ranking pipeline. When ranking code solutions using only one of the mentioned features, we can see that \textit{cluster sizes} are more important than \textit{pass rates}. However, this does not mean that \textit{pass rates} are unimportant. When both features are combined, the results can be improved even further. Finally, we achieve the best SRank performance by combining both of these features or one of them with the Interaction matrix.

It's worth noting that the cluster features possess a general and adaptable nature, allowing for potential extensions to incorporate additional information such as likelihood. We also report results of the combination of clusters' likelihood and Coder-Reviewer ranking criteria as cluster features with functional overlap in Appendix~\ref{sec:appendix-coderreviewer}.

\paragraph{Impact of Interaction Matrix}
We conduct additional experiments to demonstrate the effectiveness of the interaction matrix $\mathbf{I}$. From the results in Table \ref{tab:ablation-study}, it is obvious that the interaction matrix $\mathbf{I}$ helps to boost performance for any cluster features. Importantly, integrating \textit{cluster sizes} with $\mathbf{I}$ achieves results on par with CodeT, highlighting the significance of interactions among clusters.

\paragraph{Scaling Number of Generated Test Cases}
\begin{figure*}
\centerline{\includegraphics[width=1.1\linewidth]{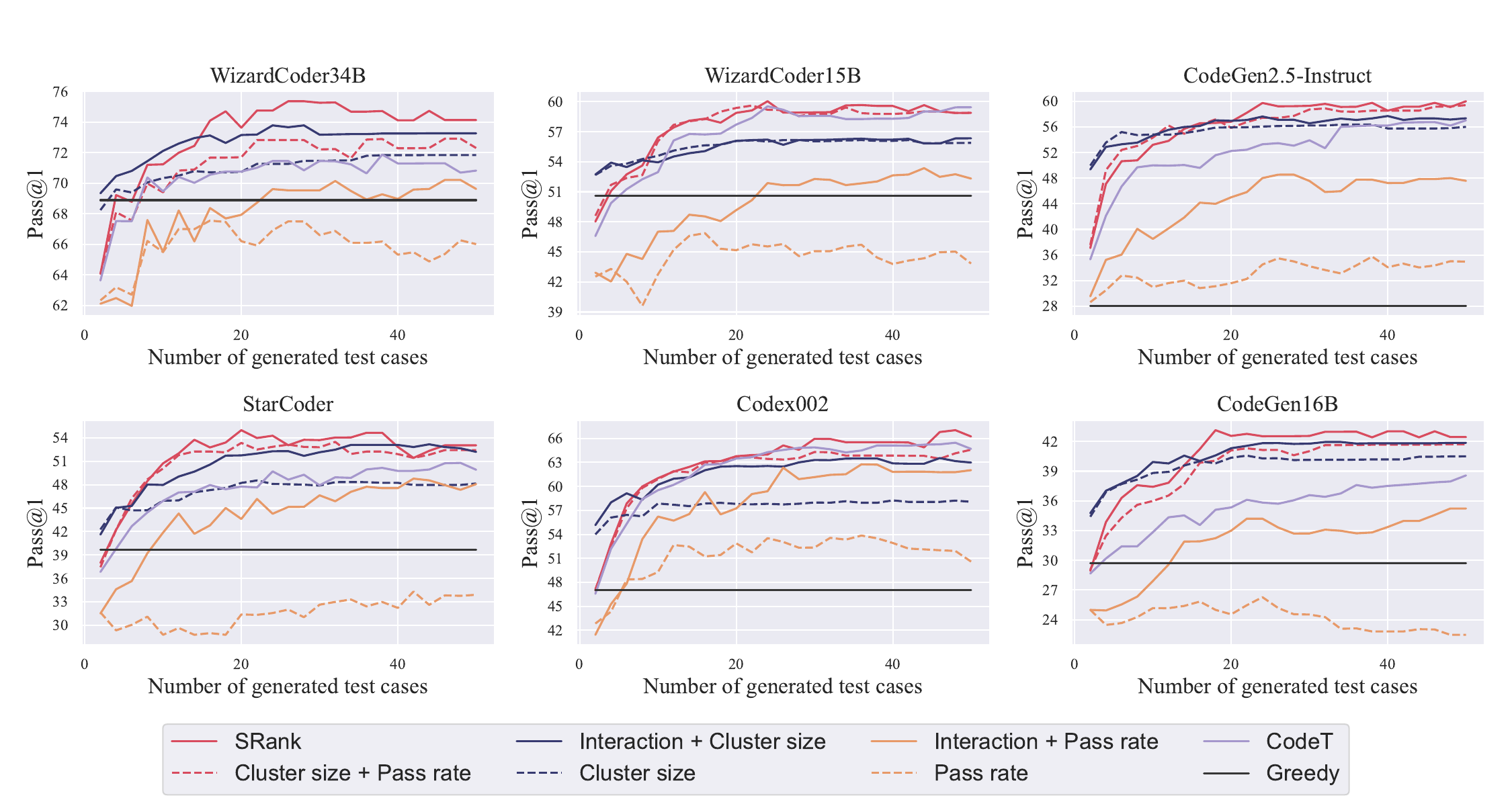}}
\caption{Ablation study on scaling number of model generated test cases vs. pass@1 on HumanEval.}\label{fig:scaling_tests_humaneval}
\vspace{-1em}
\end{figure*}
We conducted an ablation study to assess how the number of generated test cases influences code reranking performance. Figure \ref{fig:scaling_tests_humaneval} shows the pass@1 performance as a function of the number of test cases generated, ranging from 2 to 50, on the HumanEval benchmark. Please refer to Appendix~\ref{sec:appendix-ablation} for the results of MBPP-S.

Comparing each pair of solid line (with interaction matrix) and dashed line (w/o interaction matrix), cluster interaction consistently enhances performance over solely ranking by cluster features. The performance gap of reranking with and without interaction increases as the number of generated test cases increases, showcasing our method's scalability. However, with limited test cases, {\methodnamews} sometimes underperforms due to the potentially negative impact of the cluster features when integrated with cluster interaction. For an optimal balance between effectiveness and efficiency, we suggest generating at least 30 test cases to fully benefit from cluster interaction, feasible within 1 to 2 sampling rounds given many test cases are generated within a single sampled sequence.

Additionally, comparing the performance of our method and CodeT reveals disparities in some CodeLLMs. This discrepancy arises from the clustering quality of CodeT, such that limited, low-quality test cases can lead to semantically inconsistent clusters, making ranking them less meaningful. In contrast, our method clusters solutions by matching execution outputs, which ensures functional consistency, enhancing ranking reliability.

\paragraph{Scaling Number of Sampled Solutions}
In a real-world setting, it is practical to have a limit on the number of both solutions and test cases generated, since sampling multiple sequences from LLM is costly and time inefficient for low resource computing. In this section, we examine the efficiency of our {\methodnamews} by scaling down the number of sampled solutions to be less than or equal to 50 samples.

\begin{figure*}
\centerline{\includegraphics[width=1.1\linewidth]{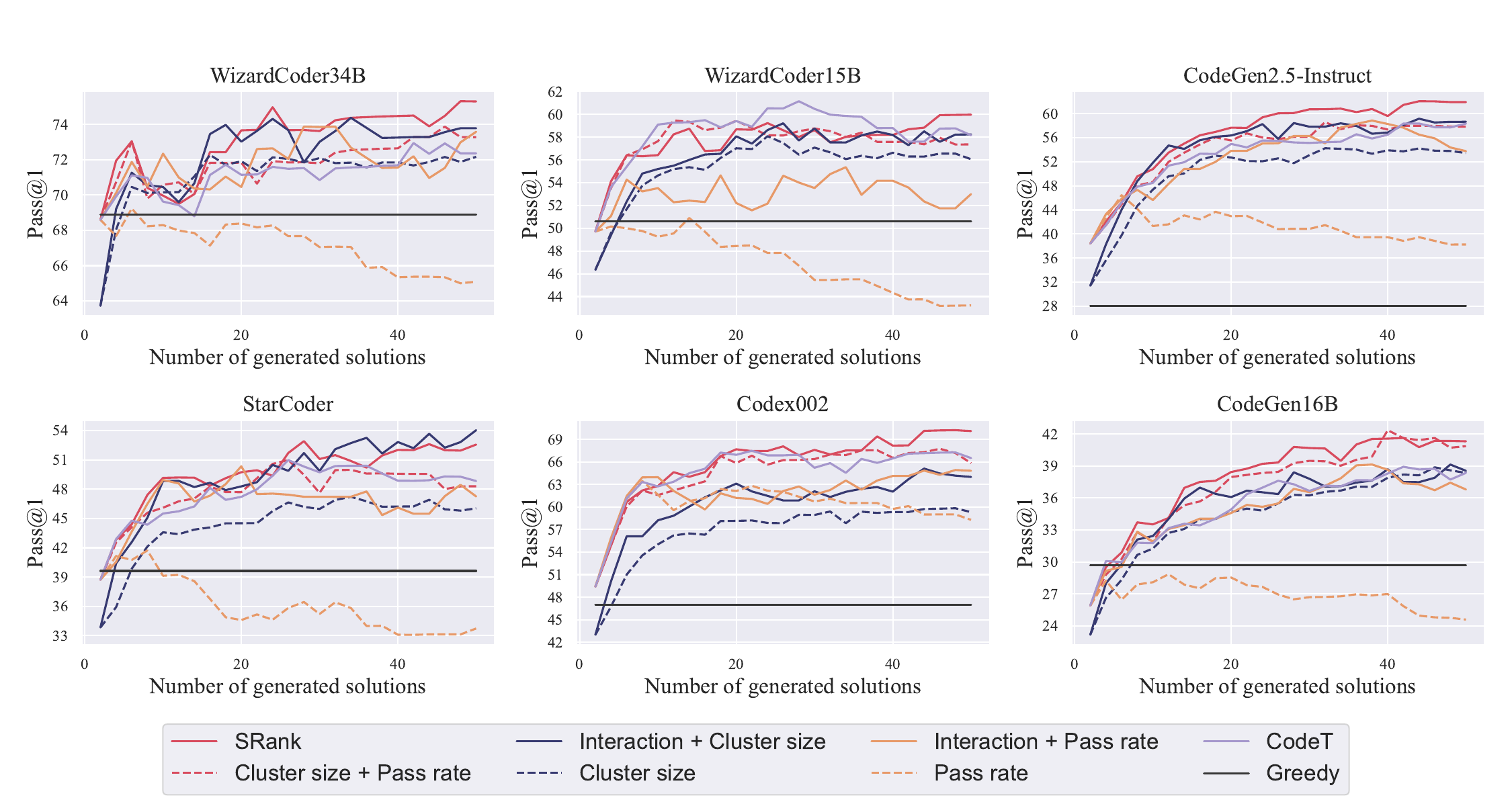}}
\caption{Ablation study on scaling number of sampled solutions vs. pass@1 on HumanEval. }\label{fig:scaling_solutions_humaneval}
\vspace{-1em}
\end{figure*}

For practical reasons, we only run each experiment with 50 test cases generated by each model, rather than all test cases extracted from 100 sequences of test cases. We prompt WizardCoder34B, WizardCoder15B, and CodeGen2.5-Instruct to generate 50 different test cases in a single sequence. This prompting technique improves the overall pipeline efficiency by significantly lowering the computational overhead of sampling from CodeLLM. Figure \ref{fig:scaling_solutions_humaneval} shows the pass@1 performance with the number of sampled solutions ranging from 2 to 50. Results of the MBPP-S benchmark can be found in Appendix~\ref{sec:appendix-ablation}. The figure shows a similar trend to when we scale the number of generated test cases. Indeed, adding cluster interaction along with cluster features brings a performance boost, even when the number of sampled code solutions is small.

Compared to CodeT, as explained earlier, due to the semantically consistent clustering process in our method, both reranking with or without interaction outperform CodeT.

Moreover, with a limited number of sampled solutions and test cases, the results bypass greedy search (represented by the black lines). This proves both the effectiveness and the efficiency of our approach.

\section{Related Work}
\paragraph{Code Large Language Models}
The emergence of large language models (LLMs) has transformed code understanding and generation tasks \citep{rozi2023codellama, li2023starcoder, nijkamp2023codegen2, wang2023codet5plus, nijkamp2023codegen, fried2023incoder, li2023textbooks}. Recent research has employed natural language processing models for code-related tasks, using pretraining strategies akin to those for natural languages \citep{codebert, codet5, guo2020graphcodebert, ahmad2021unified, elnaggar2021codetrans, peng2021could, kanade2020learning, chakraborty2022natgen, ahmed2022multilingual, niu2022spt}. Among various CodeLLMs, those with larger capacities tend to perform better in code generation tasks. For instance, StarCoder \citep{li2023starcoder} and CodeLlama \citep{rozi2023codellama} handle contexts with up to 8,000 and 100,000 tokens, respectively, while WizardCoder \citep{luo2023wizardcoder} excels in evolving instructions. Additionally, models like CodeGen2.5-Instruct \citep{nijkamp2023codegen2}, CodeT5+Instruct \citep{wang2023codet5plus}, Codex002 \citep{chen2021_eval_llms}, CodeGen-Mono 16B \citep{nijkamp2023codegen}, and InCoder 6B \citep{fried2023incoder} have also shown promise in this domain.

\paragraph{Reranking Methods for Code Generation}
Several studies have explored reranking code generated by language models \citep{chen2021_eval_llms, zhang2023coderreviewer, ni2023lever, chen2023codet, inala2022faultaware, to10better}, prioritizing solutions from these models. \citet{chen2021_eval_llms} showed empirically that selecting solutions based on the mean log probability of tokens improves performance. Coder-Reviewer \citep{zhang2023coderreviewer} proposed a mutual information-based ranking method for natural language instructions and generated solutions. Reranking has also been approached using execution-based metrics. MBR-exec \citep{shi2022mbr_exec} minimizes a loss function across all solutions, while AlphaCode \citep{li2022competition} clusters solutions based on execution outputs. LEVER \citep{ni2023lever} uses a verifier to assess program correctness, and CodeT \citep{chen2023codet} generates high-quality test cases. Our approach stands out as it does not require model training or fine-tuning and can complement methods like LEVER \citep{ni2023lever}.

\section{Conclusion}
We propose {\methodnamews}, a novel reranking strategy designed to extract optimal code generation solutions from CodeLLMs. {\methodnamews} focuses on modeling inter-cluster relationships to identify clusters with the highest functional overlap with others. By prioritizing the cluster with the most comprehensive interaction, often indicating correct functionality, we can pinpoint the most promising solution. Incorporating these inter-cluster relationships into the ranking pipeline enhances solution identification.

We showcase the state-of-the-art performance of \textbf{\methodnamews} on pass@1 across various well-known CodeLLMs, surpassing other ranking methods like CodeT and Coder-Reviewer in extensive evaluations. Our thorough analysis further highlights the method's effectiveness in realistic scenarios with a limited number of solutions and test cases. These findings are crucial as they address the challenges of code generation in real-world applications, illuminating strategies for selecting superior solutions within constrained coding environments.
\section*{Limitations}
Our approach presents several opportunities for further development and refinement. While our method demonstrates superiority over others in our empirical evaluation, which focuses on Python, extending {\methodnamews} to a multilingual benchmark would provide a more comprehensive assessment of its effectiveness across different programming languages.


{\methodnamews} has shown promising results in its current scope, which primarily focuses on functions with clear output metrics, such as arithmetic or algorithmic problems. To extend {\methodnamews} to more complex coding tasks, such as system architecture, memory management, and process management, we would need to validate that our underlying assumption of diverse incorrect solutions holds true in these domains and adapt the function $\delta$ to effectively capture the intricacies of these tasks. This presents an exciting opportunity to expand the applicability of our approach.

While implementing {\methodnamews} could introduce computational overhead due to the need to generate a large number of candidate solutions and test cases, our ablation study demonstrates that {\methodnamews} can achieve superior performance with a limited number of sampled solutions and test cases. To further optimize the efficiency of our method, particularly in resource-constrained environments or when rapid solution generation is crucial, we plan to conduct additional analyses and explore potential optimizations. Currently, {\methodnamews} focuses on assessing functional correctness by modeling the functional overlap among solutions. While functional correctness is a critical aspect of code quality, we acknowledge the importance of other factors such as efficiency and robustness. Integrating these factors into our framework presents an exciting direction for future work, as it would enhance the comprehensiveness and robustness of {\methodnamews} in meeting the diverse requirements of real-world coding tasks. By addressing these aspects, we aim to make {\methodnamews} an even more valuable tool for the coding community.

\bibliography{anthology,custom}
\bibliographystyle{acl_natbib}

\appendix
\section{Validity of our assumption}
\label{sec:appendix-valid-assumption}
\subsection{Quantitative Assessment}

To consolidate our assumption, we provide additional results besides Section \ref{sec:analysis}. Formally, we introduce additional notations. $s_i = s_j$ is defined as the two solutions sharing the same semantic functionality, while $s_i \neq s_j$ denotes the opposite. We then compute the probability of having two equivalently semantic solutions below:
\begin{equation}
p(s_i=s_j) = \frac{\sum_{k}\binom{|C_k|}{2}}{\binom{|F|}{2}}
\end{equation}
Given that our clustering algorithm ensures all the solutions within a cluster share functionality, and solutions from distinct clusters are certainly not equivalently semantic, we just need to randomly select two solutions from a cluster to satisfy $s_i=s_j$. Thus, the probability of having two equivalently semantic incorrect solutions is
\begin{equation}
\begin{split}
p(s_i=s_j,  s_i~\text{and} ~s_j~ \text{are incorrect}) \\ 
= \frac{\sum_{k}\binom{|C_k|^{*}}{2}}{\binom{|F|}{2}}
\end{split}
\end{equation}
We can easily compute the probability of two wrong solutions given that they are equivalently-semantic
\begin{equation}
\begin{split}
p(s_i~\text{and} ~s_j~ \text{are incorrect}|s_i=s_j) \\=\frac{p(s_i=s_j,  s_i~\text{and} ~s_j~ \text{are incorrect}) }{p(s_i=s_j) }
\end{split}
\end{equation}
Results in Table \ref{tab:prob-match-pairs} show that the probability of having two equivalently semantic incorrect solutions is pretty low at around 0.092. Given that two solutions share the same functionality, the probability for them to be wrong is approximately 0.262, far lower than that of having them be correct. These findings provide strong support for our assumption.


\begin{table*}
\centering
\begin{tabular}{lccc}
\toprule
Model               & $p(s_i = s_j)$ & $p(s_i = s_j, s_i~\text{and} ~s_j~ \text{are incorrect})$ & $p( s_i~\text{and} ~s_j~ \text{are incorrect} | s_i = s_j)$ \\ \midrule
StarCoder           & 0.2176         & 0.0559      & 0.2569                                             \\
CodeGen2.5-Instruct & 0.2692         & 0.0737       &0.2738                                            \\
WizardCoder15B      & 0.4596         & 0.1134      &0.2467                                             \\
WizardCoder34B      & 0.5689         & 0.0990             &0.1740                                      \\
Codex002            & 0.3218         & 0.0807    &   0.2508                                            \\
CodeGen16B          & 0.3482         & 0.1279   & 0.3673                                               \\
\bottomrule
\end{tabular}
\caption{Probabilities of two equivalently-semantic solutions and two equivalently-semantic incorrect solutions among different models}
\label{tab:prob-match-pairs}
\end{table*}

\subsection{Qualitative Assessment}
\begin{figure}
  \begin{subfigure}{0.49\linewidth} 
    \centering
    \includegraphics[width=\linewidth]{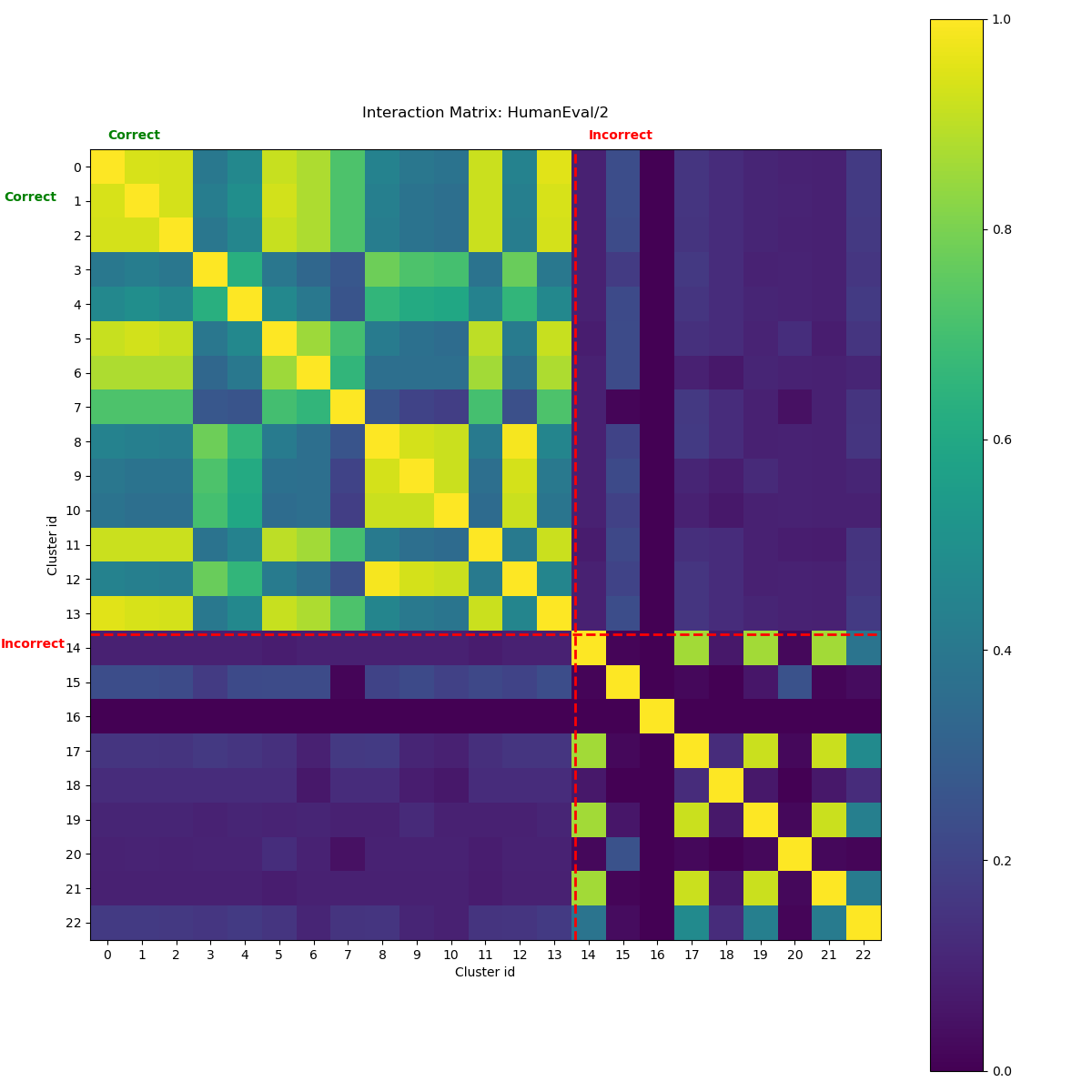} 
  \end{subfigure}%
  \begin{subfigure}{0.49\linewidth} 
    \centering
    \includegraphics[width=\linewidth]{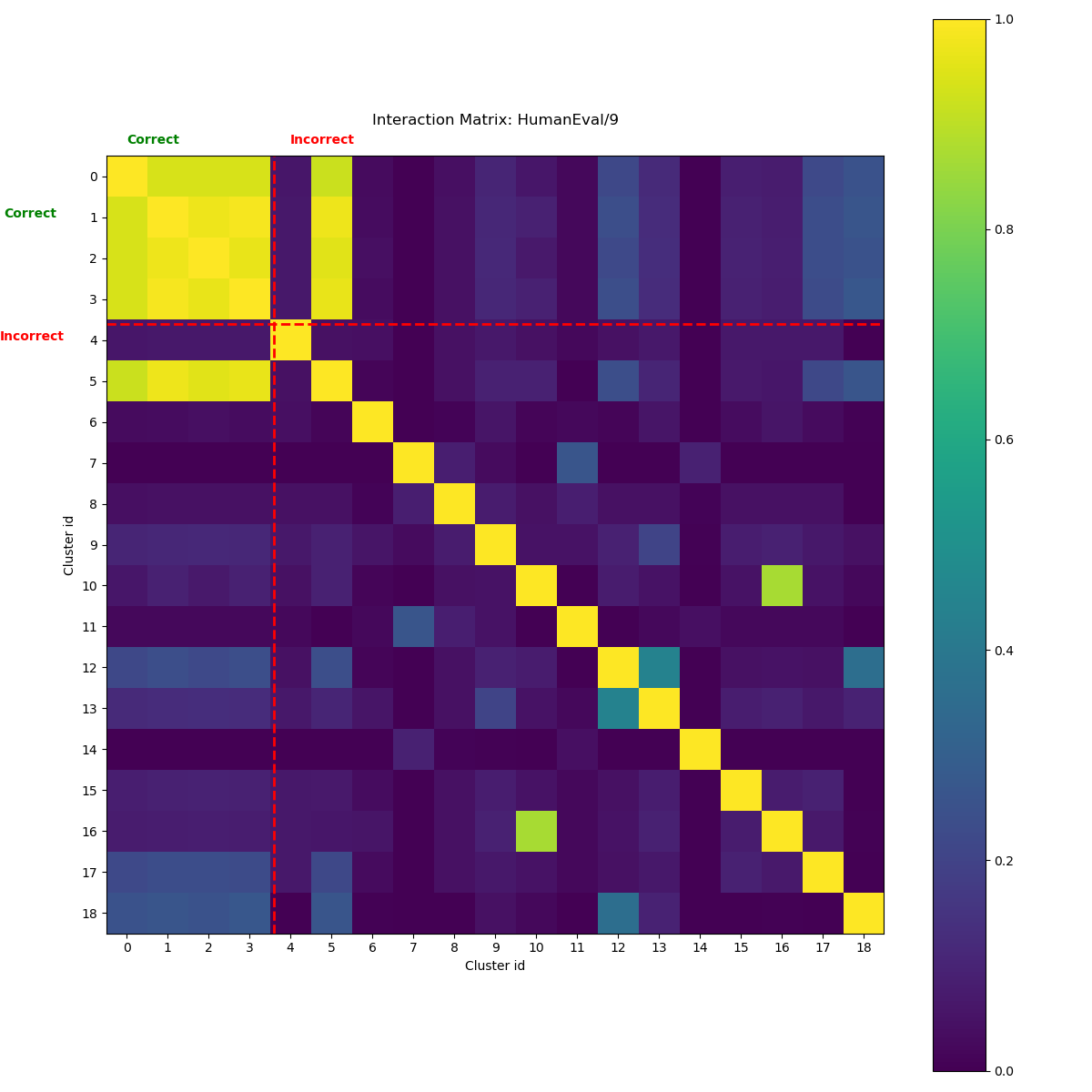} 
  \end{subfigure}
  \caption{Demonstration of our assumption shows there is a low functional agreement among incorrect solutions. The two graphs represent two problems chosen from Human-Eval, with solutions generated by StarCoder.}
  \label{fig:demo-matrix-assump}
\end{figure}

Besides quantitative evaluation supporting our assumption, we offer qualitative examples through interaction matrices of solutions generated by StarCoder on HumanEval, as shown in Figure \ref{fig:demo-matrix-assump}. Each matrix is divided into four separate patches by two red-dashed lines: top-left, top-right, bottom-left, and bottom-right. The left region of the vertical line represents clusters, including correct solutions, while the right region encompasses incorrect clusters; similarly, the top region above the horizontal line comprises correct solutions. The diagonal lines are disregarded as their values represent the self-interaction of clusters. It is clearly seen that the top-left patches are notably brighter, whereas the bottom-right patches are virtually dark. These observations demonstrate that correct clusters interact with each other comprehensively, indicating a high probability of functional overlap between correct solutions, while the opposite holds true for incorrect solutions.

\section{Generation Prompts}
\label{sec:appendix-prompt}
In this section, we list all the prompts feeding to CodeLLMs for generation process of code solutions and test cases. For concrete demonstration, we use the same problem `HumanEval/8` throughout this section.
\subsection{Code generation}
Demonstrated prompts for code generation includes Figure~\ref{fig:prompt_code_gen_1st}, Figure ~\ref{fig:prompt_code_gen_2nd}, Figure~\ref{fig:prompt_code_gen_3rd}, and Figure~\ref{fig:prompt_code_gen_4th}.

\begin{figure}
\begin{lstlisting}
Below is an instruction that describes a task. Write a response that approriately completes the requests.

### Instruction:
Create a Python script for this problem:
from typing import List, Tuple

def sum_product(numbers: List[int]) -> Tuple[int, int]:
    """ For a given list of integers, return a tuple consisting of a sum and a product of all the integers in a list. Empty sum should be equal to 0 and empty product should be equal to 1."""

### Response:
\end{lstlisting}
\caption{Code generation prompt used with WizardCoder34B, WizardCoder15B, and CodeGen2.5-Instruct}
\label{fig:prompt_code_gen_1st}
\end{figure}
\begin{figure}
\begin{lstlisting}<filename>solutions/solution_1.py
# Here is the correct implementation of the code exercise
from typing import List, Tuple

def sum_product(numbers: List[int]) -> Tuple[int, int]:
    """ For a given list of integers, return a tuple consisting of a sum and a product of all the integers in a list. Empty sum should be equal to 0 and empty product should be equal to 1."""
\end{lstlisting}
\caption{Code generation prompt used with StarCoder}
\label{fig:prompt_code_gen_2nd}
\end{figure}

\begin{figure}
\begin{lstlisting}
from typing import List, Tuple

def sum_product(numbers: List[int]) -> Tuple[int, int]:
    """ For a given list of integers, return a tuple consisting of a sum and a product of all the integers in a list. Empty sum should be equal to 0 and empty product should be equal to 1."""
\end{lstlisting}
\caption{Code generation prompt used with Codex002, CodeGen16B}
\label{fig:prompt_code_gen_3rd}
\end{figure}

\begin{figure}
\begin{lstlisting}
Complete the following Python script. We will be using the output you provide as-is to create new files, so please be precise and do not include any other text. Your output needs to be ONE file; Otherwise, it will break the system. Moreover, your response must include the provided code and the newly generated code. Your output must consist ONLY of the language and code, in the fenced code block format:
```language
CODE
```
Here is the initial code:
```python
from typing import List, Tuple

def sum_product(numbers: List[int]) -> Tuple[int, int]:
    """ For a given list of integers, return a tuple consisting of a sum and a product of all the integers in a list. Empty sum should be equal to 0 and empty product should be equal to 1."""
```
\end{lstlisting}
\caption{Code generation prompt used with Claude 3 Opus}
\label{fig:prompt_code_gen_4th}
\end{figure}

\subsection{Test Generation}
We give example test case generation prompts in Figure~\ref{fig:prompt_test_gen_1st}, Figure ~\ref{fig:prompt_test_gen_2nd}, Figure~\ref{fig:prompt_test_gen_3rd}, and Figure~\ref{fig:prompt_test_gen_4th}.
\begin{figure}
\begin{lstlisting}
Below is an instruction that describes a task. Write a response that appropriately completes the request.

### Instruction:
I have this function stub, please generate 50 test cases for this function. The function stub is as follow:
```python
from typing import List, Tuple

def sum_product(numbers: List[int]) -> Tuple[int, int]:
    """ For a given list of integers, return a tuple consisting of a sum and a product of all the integers in a list. Empty sum should be equal to 0 and empty product should be equal to 1."""
    pass
```
- Each test case is in the form of assertion statement, for example: assert sum_product(...) == ...
- Each test case is in a single line 
- The length of each test case should be too long, ideally less than or equal to 150 letters
- The test input should not be too long
- The inputs of test cases should be diverse and cover corner cases of the function
- Test cases should not be repeated

### Response: Here are 50 test cases for function `sum_product`:
assert sum_product(
\end{lstlisting}
\caption{Test case generation prompt used with WizardCoder34B, WizardCoder15B, and CodeGen2.5-Instruct}
\label{fig:prompt_test_gen_1st}
\end{figure}

\begin{figure}
\begin{lstlisting}
<filename>solutions/solution_1.py
from typing import List, Tuple

def sum_product(numbers: List[int]) -> Tuple[int, int]:
    """ For a given list of integers, return a tuple consisting of a sum and a product of all the integers in a list. Empty sum should be equal to 0 and empty product should be equal to 1."""
    pass

# check the correctness of sum_product
assert sum_product(
\end{lstlisting}
\caption{Test case generation prompt used with StarCoder}
\label{fig:prompt_test_gen_2nd}
\end{figure}

\begin{figure}
\begin{lstlisting}from typing import List, Tuple

def sum_product(numbers: List[int]) -> Tuple[int, int]:
    """ For a given list of integers, return a tuple consisting of a sum and a product of all the integers in a list. Empty sum should be equal to 0 and empty product should be equal to 1."""
    pass

# check the correctness of sum_product
assert sum_product(
\end{lstlisting}
\caption{Test generation prompt used with Codex002, CodeGen16B}
\label{fig:prompt_test_gen_3rd}
\end{figure}

\begin{figure}
\begin{lstlisting}
You are given a function stub. Your task is to generate 20 test cases for this function. Each test case must be in the form of a single-line assertion statement in Python. For example: `assert function_name(input) == output`. Replace `function_name` with the actual function name from the stub, and replace `input` and `output` with the appropriate values. The test inputs must be diverse and cover all possible edge cases of the function. Do not repeat any test case. Additionally, each test case input must not be too computationally expensive when executing the function. Your output must consist ONLY of the test cases, with each test case on a new line. Do not include any other text.

The function stub is as follows:
```python
from typing import List, Tuple

def sum_product(numbers: List[int]) -> Tuple[int, int]:
    """ For a given list of integers, return a tuple consisting of a sum and a product of all the integers in a list. Empty sum should be equal to 0 and empty product should be equal to 1."""
```
\end{lstlisting}
\caption{Test generation prompt used with Claude 3 Opus}
\label{fig:prompt_test_gen_4th}
\end{figure}
\subsection{CoderReviewer Prompts}
In CoderReviewer, Coder generates code solution given the task description and Reviewer checks the generated solution by measuring the sequence probability of the description given the solution.

For generating code solutions, we use the same prompt as in the Code Generation section. For Reviewer, to adapt to instruction-tuned models, here we list the prompt we manually design to calculate the probability of task description given the code implementation with respect to CodeLLMs.

The prompt templates we used including Figure~\ref{fig:prompt_coderreviewer_1st}, Figure ~\ref{fig:prompt_coderreviewer_2nd}, and Figure~\ref{fig:prompt_coderreviewer_3rd}.

\begin{figure}
\begin{lstlisting}Below is an instruction that describes a task. Write a response that appropriately completes the request.

### Instruction:
Write a docstring for the above function:
from typing import List, Tuple

def sum_product(numbers: List[int]) -> Tuple[int, int]:
    product = 1
    for number in numbers:
        product *= number
    return (sum(numbers), product)

### Response: Here's the docstring for the above function
def sum_product(numbers: List[int]) -> Tuple[int, int]:
    """ For a given list of integers, return a tuple consisting of a sum and a product of all the integers in a list. Empty sum should be equal to 0 and empty product should be equal to 1."""
\end{lstlisting}
\caption{Prompt used for measuring sequence probability of task description given a specific code implementation of models WizardCoder34B, WizardCoder15B, and CodeGen2.5-Instruct.}
\label{fig:prompt_coderreviewer_1st}
\end{figure}

\begin{figure}
\begin{lstlisting}<filename>solutions/solution_1.py
from typing import List, Tuple

def sum_product(numbers: List[int]) -> Tuple[int, int]:
    product = 1
    for number in numbers:
        product *= number
    return (sum(numbers), product)

# Write a docstring for the above function
def sum_product(numbers: List[int]) -> Tuple[int, int]:
    """ For a given list of integers, return a tuple consisting of a sum and a product of all the integers in a list. Empty sum should be equal to 0 and empty product should be equal to 1."""

\end{lstlisting}
\caption{Prompt used for measuring sequence probability of task description given a specific code implementation of model StarCoder}
\label{fig:prompt_coderreviewer_2nd}
\end{figure}

\begin{figure}
\begin{lstlisting}
from typing import List, Tuple

def sum_product(numbers: List[int]) -> Tuple[int, int]:
    product = 1
    for number in numbers:
        product *= number
    return (sum(numbers), product)

# write a docstring for the above function
def sum_product(numbers: List[int]) -> Tuple[int, int]:
    """ For a given list of integers, return a tuple consisting of a sum and a product of all the integers in a list. Empty sum should be equal to 0 and empty product should be equal to 1."""
\end{lstlisting}
\caption{Prompt used for measuring sequence probability of task description given a specific code implementation of models Codex002, CodeGen16B}
\label{fig:prompt_coderreviewer_3rd}
\end{figure}

\section{Scaling Number of Generated Test Cases and Sampled Solutions on MBPP-S}
\label{sec:appendix-ablation}
\begin{figure*}
\centerline{\includegraphics[width=\linewidth]{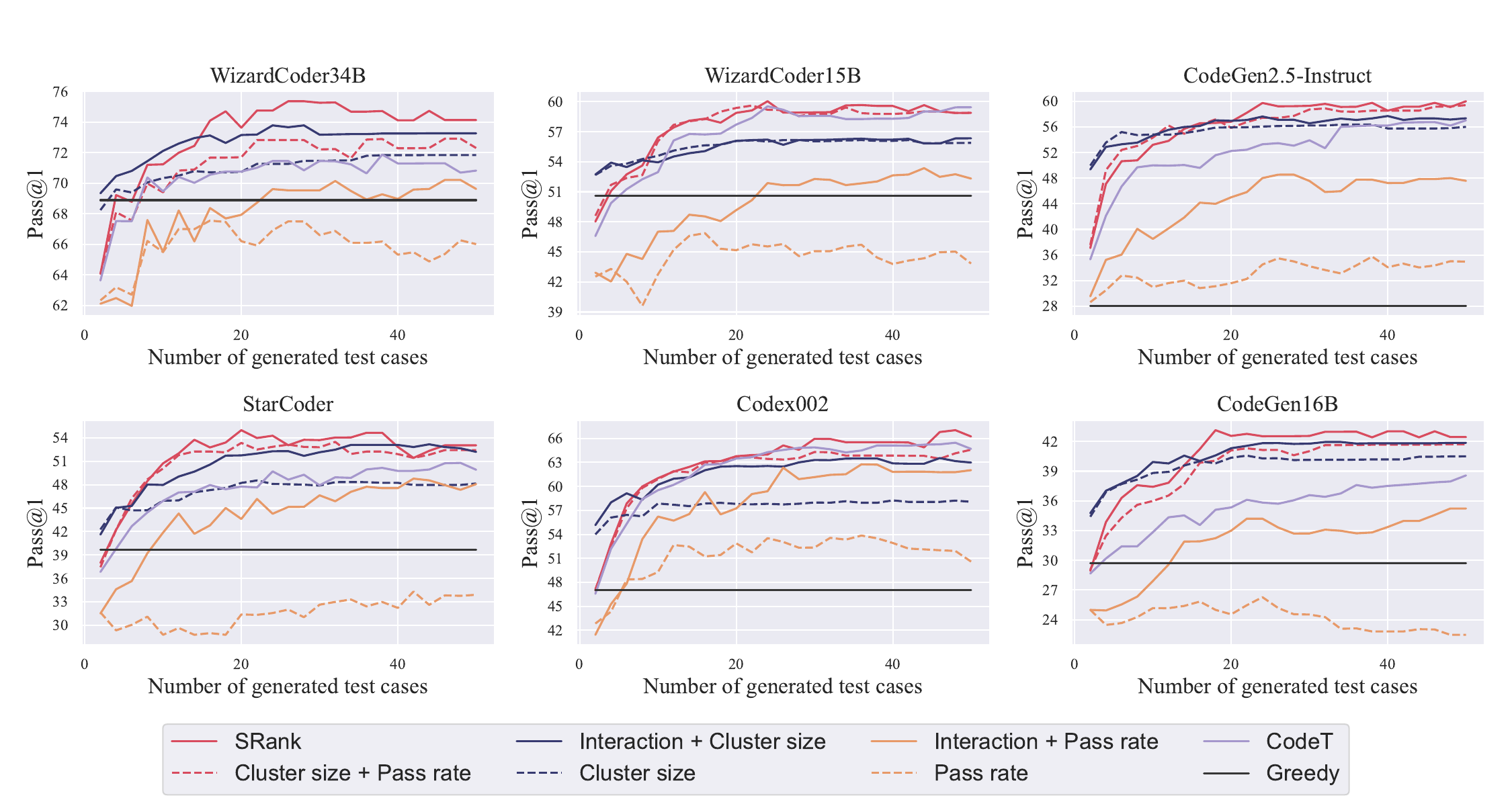}}
\caption{Ablation study on scaling number of model generated test cases vs. pass@1 on MBPP-S.}\label{fig:scaling_tests_mbpp}
\vspace{-1em}
\end{figure*}
\begin{figure*}
\centerline{\includegraphics[width=\linewidth]{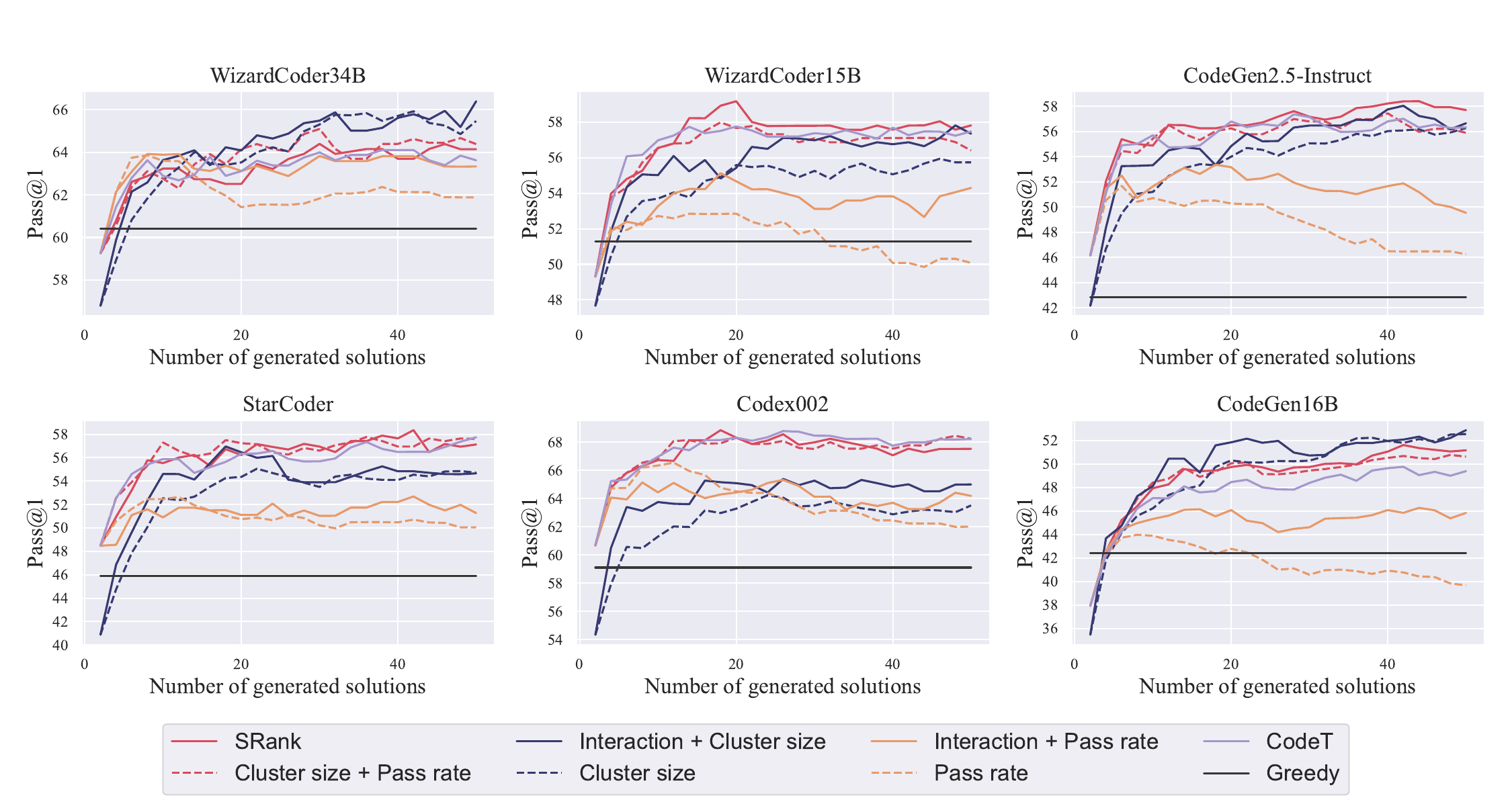}}
\caption{Ablation study on scaling number of sampled solutions vs. pass@1 on MBPP-S.}\label{fig:scaling_solutions_mbpp}
\vspace{-1em}
\end{figure*}
This section complements the results on MBPP-S dataset from Section Scaling Number of Generated Test Cases and Section Scaling Number of Sampled Solutions. The setting for MBPP-S is the same as HumanEval mentioned in each above section. Figure~\ref{fig:scaling_tests_mbpp} and Figure~\ref{fig:scaling_solutions_mbpp} shows pass@1 as function of the number of generated test cases and sampled solutions respectively on MBPP-S benchmark.

\section{CoderReviewer Ranking Criteria as Cluster Features}
\label{sec:appendix-coderreviewer}
\begin{table}[t]
\begin{center}
    \scalebox{0.85}{
\begin{tabular}{lccc}
\toprule
{} &            HumanEval &           MBPP-S \\
\midrule
N.Coder            &              $57.15$ &              $56.73$ \\
Interaction + N.Coder            &              $62.69$ &              $59.50$ \\
\midrule
N.Reviewer            &              $55.33$ &              $55.41$ \\
Interaction + N.Reviewer            &              $62.69$ &              $59.74$ \\
\midrule
N.CoderReviewer            &              $62.67$ &              $59.39$ \\
Interaction + N.CoderReviewer            &              $\mathbf{63.30}$ &              $\mathbf{61.38}$ \\
\bottomrule
\end{tabular}
    }
    \caption{CoderReviewer ranking criterions as cluster fearures.}
    \label{tab:coderreviewer}
\end{center}
\end{table}
We investigate adding cluster interaction with CoderReviewer ranking criteria as cluster features, where Coder represents the likelihood $P_{CodeLLM}(x|y)$ and Reviewer represents $P_{CodeLLM}(y|x)$. CoderReviewer is the multiplication $P_{CodeLLM}(x|y)P_{CodeLLM}(x|y)$, with $x$ as the task description and $y$ as generated code solution.
The prefixed N. before each term indicates that the likelihood is normalized by the number of sequence tokens, which was shown to improve performance in previous work. 
The cluster feature in these cases is the criteria score of the solution with maximum score within that cluster. 

Experimental results in Table \ref{tab:coderreviewer} show consistent improvement when adding cluster interaction over CoderReviewer. This demonstrates that when the interaction matrix is combined with CoderReviewer, the interaction matrix can improve CoderReviewer's performance even further, indicating that our method is adaptable to different reranking methods.
\section{Reranking with Interaction Only}
\label{sec:appendix-interaction-only}
\begin{table*}[t]
\begin{center}
	\scalebox{0.7}{
		\begin{tabular}{lcccccc}
			\toprule
			{} & \multicolumn{6}{c}{\textbf{HumanEval}} \\
			\cmidrule(lr){2-7}
			{} & WizardCoder34B & WizardCoder15B & CodeGen2.5-Instruct & StarCoder & Codex002 & CodeGen16B \\
			\midrule
			Greedy            & \textbf{68.90} & \textbf{50.61} & 28.05 & 39.63 & 47.00 & 29.70 \\
			Random            & 59.88 & 45.20 & 26.68 & 32.55 & 37.06 & 22.78 \\
            \midrule
			Interaction Only  & 66.49 & 49.79 & \textbf{51.13} & \textbf{50.01} & \textbf{62.75} & \textbf{31.31} \\
			\midrule
			
			{} & \multicolumn{6}{c}{\textbf{MBPP-S}} \\
			\cmidrule(lr){2-7}
			{} & WizardCoder34B & WizardCoder15B & CodeGen2.5-Instruct & StarCoder & Codex002 & CodeGen16B \\
			\midrule
			Greedy            & \textbf{60.42} &51.29 & 42.86 & 45.90 & \textbf{58.10} & \textbf{42.40} \\			
			Random            & 54.37 & 45.72 & 34.60 & 39.26 & 47.50 & 31.54 \\
            \midrule
            Interaction Only & 59.60 & \textbf{53.94} & \textbf{47.70} & \textbf{48.85} & 57.49 & 41.98 \\
			\bottomrule
		\end{tabular}
	}
 \captionsetup{type=table}
	\caption{Results of pass@1 on HumanEval and MBPP-S benchmarks when reranking merely by interaction among cluster without any cluster feature.}
	\label{tab:interaction-only}
\end{center}
\end{table*}
To further validate that modeling inter-cluster interactions aids in ranking clusters, we present the results of reranking solely by matrix I without any cluster features V in Table~\ref{tab:interaction-only}. In this scenario, V is a column vector with 1 at every entry. Consequently, the ith entry in R is the sum of functional overlap between the ith cluster and all other clusters.

The results demonstrate that incorporating interaction modeling consistently elevated performance beyond random. Furthermore, reranking with interaction alone significantly outperforms the performance of greedy decoding in some cases, such as 51.13 vs 28.49 for CodeGen2.5, 50.01 vs 39.63 for StarCoder, 62.75 vs 47.00 for CodeX002 in HumanEval. This result futher confirms the effectiveness of our method \textbf{\methodnamews}.

\section{Additional Results with Closed-Source LLM}
\label{sec:closed-source-results}
\begin{table}[t]
\begin{center}
	\scalebox{0.7}{
 		\begin{tabular}{lc}
			\toprule
			Method & Claude 3 Opus \\
			\midrule
   			Random            & 78.03 \\
			Greedy            & 78.05 \\
			CodeT             & 77.42 \\
            \midrule
			SRank  & \textbf{79.18} \\
			\bottomrule
		\end{tabular}
	}
 \captionsetup{type=table}
	\caption{Results of pass@1 on HumanEval in the zero-shot setting of closed-source model Claude 3 Opus with different LLM decoding strategies and reranking methods.}
	\label{tab:closed-source-results}
\end{center}
\end{table}
In the previous section, we presented code generation ranking results with both open-source and closed-source models (Codex002), which is not the strongest and up-to-date model for code. To examine our method on contemporary and high-capability LLMs, we selected Anthropic Claude 3 Opus \citep{anthropic2024claude3} as a representative model for our experiments. Table~\ref{tab:closed-source-results} shows the results of code generation on HumanEval using different decoding strategies, such as greedy search and random sampling, as well as when reranking methods like {\methodnamews} and CodeT are employed. Even with such a large and powerful LLM, {\methodnamews} still achieves impressive performance compared to other baselines, while CodeT degrades performance of baselines.

In this experiment, we followed similar hyperparameter settings as with other models, setting the temperature to 0.8 and top-p to 0.95. The results indicate that random sampling and greedy search yield roughly the same performance, suggesting that the model is very confident in its responses and the randomness in output sequences is minimal. This behavior is further confirmed by observing the generated code solutions and test cases, where there is little variance between the samples and a higher rate of duplication compared to other models we experimented with. For such large and high-capability models, the benefits from reranking strategies like {\methodnamews} are most fully realized when sufficient randomness is introduced in the sampling process by adjusting hyperparameters, such as temperature, top-k, and top-p.

Notably, the highest reported score for Claude 3 Opus from Anthropic on HumanEval is 84.9, which is higher than the numbers presented in this work. The explanation for this discrepancy is that the test cases included in the HumanEval prompt were excluded in our settings, following previous work in code generation reranking.

\section{Case Study}
\label{sec:case-study}
\begin{figure*}
    \centering
    \hspace*{-0.25cm}
    \includegraphics[width=1 \linewidth]{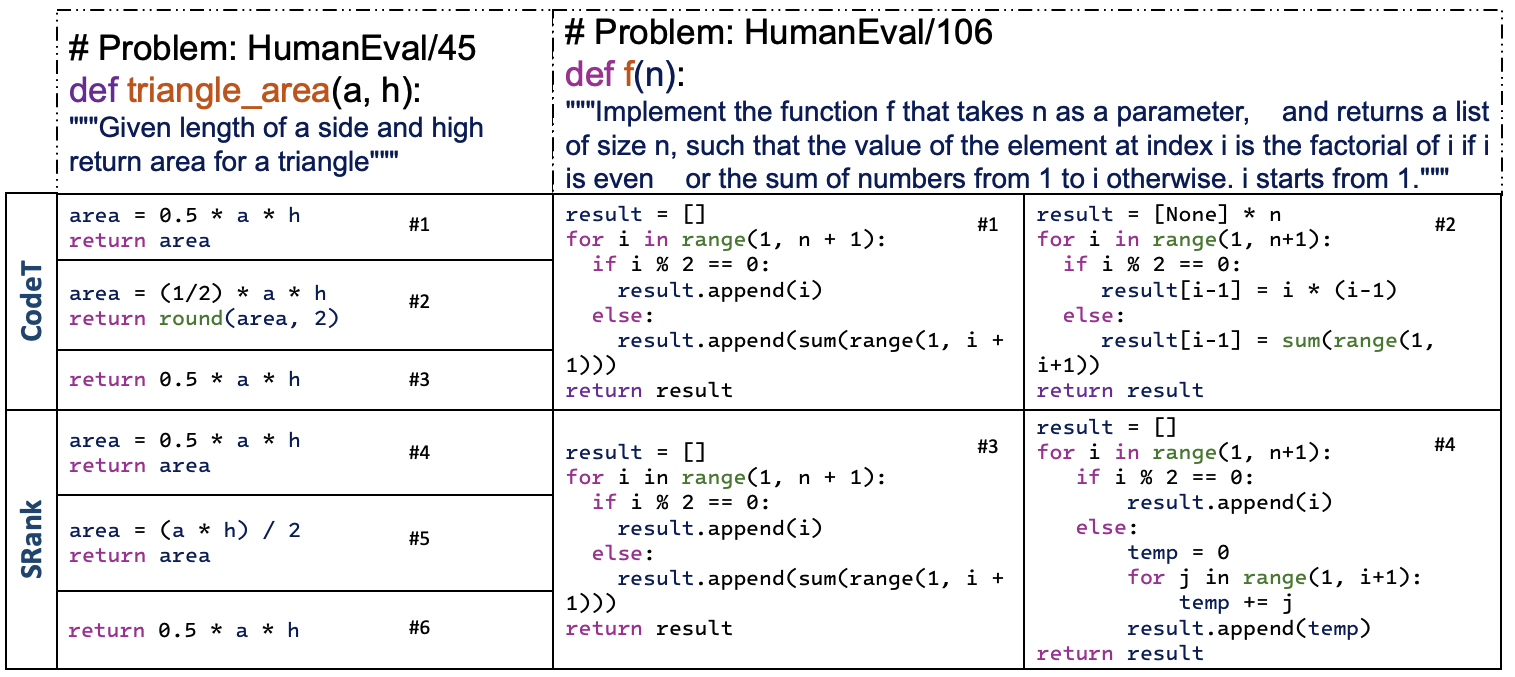}
    \caption{Case studies from HumanEval's problems with the highest-score clusters produced by CodeT versus SRank using CodeGen2.5-Instruct.}
    \label{fig:case-study}
\end{figure*}

In this section, we compare the quality of cluster samples from CodeT and our method. The top 1 clusters ranked by CodeT and SRank are shown in Figure \ref{fig:case-study}. It is clear that functional inconsistency among solutions plagues CodeT's clusters. For example, in the problem HumanEval 45, solution \#2 is not semantically equivalent to solutions \#1 and \#3, and in the problem HumanEval 106, solution \#1 differs from solution \#2. This phenomenon is explained by the fact that CodeT groups solutions that pass the same set of test cases into clusters. As a result, when computing the pass@k, inconsistency in functionalities among top cluster solutions can degrade performance. \textit{SRank}, on the other hand, considers execution outputs for the clustering process, ensuring functional consistency among solutions within a cluster.
\end{document}